\pdfoutput=1

\documentclass[aps,preprintnumbers,amsmath,amssymb,twocolumn, tightenlines,superscriptaddress,nofootinbib]{revtex4}
\usepackage{graphicx}
\usepackage{slashed}
\usepackage{tikz,mathpazo}
\usetikzlibrary{shapes.geometric, arrows}

\begin{document}


\title{Relativistic Borromean States}

\author {Ziyue Wang}
\address{Physics Department, Tsinghua University, , Beijing 100084, China.}
\author {Shao-Jian Jiang}
\address{State Key Laboratory of Magnetic Resonance and Atomic and Molecular Physics, Innovation Academy for Precision Measurement Science and Technology, Chinese Academy of Sciences, Wuhan 430071, China}
\author {Yin Jiang} 
\address{Physics Department, Beihang University, 37 Xueyuan Rd, Beijing 100191, China}
\date{\today}

\begin{abstract}
In this work the existence of Borromean states has been discussed for bosonic and fermionic cases in both the relativistic and
non-relativistic limits from the 3-momentum shell renormalization.
With the linear bosonic model we checked the existence of Efimov-like states in the bosonic system.
In both limits a geometric series of singularities are found in the 3-boson interaction vertex, while the energy ratio is reduced
by around 70\% in the relativistic limit because of the anti-particle contribution. Motivated by the quark-diquark model in
heavy baryon studies, we have carefully examined the p-wave quark-diquark interaction and found an isolated Borromean pole at finite
energy scale. This may indicate a special baryonic state of light quarks in high energy quark matters. In other cases trivial
results are obtained as expected. In relativistic limit, for both bosonic and fermionic cases, potential Borromean states are independent
of the mass, which means the results would be valid even in zero-mass limit as well.
\end{abstract}

\maketitle

{\it Introduction.---}
In classical physics, the three-body problem in gravitation systems could be traced back to Galileo Galilei.
Its nonlinearity led to a spectacular direction called dynamical system founded by Henri Poincar{\'{e}}'s works.
Except some special and stable solutions\cite{Moore, Suvakov}, a generic dynamical system is usually too complicated to solve without supercomputers.
Ineluctably this situation passed on to its descendants when married with quantum mechanics.
Nevertheless, there is also supposed to be some special solutions in quantum systems as well, which are usually
closely related to low energy or long range properties of the system\cite{Naidon}. Efimov effect, which is discovered in
the quantum spectrum of a three-heavy-particle system is one of these special solutions. Intuitively the three-body states are also referred to
as Borromean-ring states in which any pairs of these three are bound infinitely loosely.
The geometric series of three-body bound states(Efimov series), was first found in nuclear systems by Efimov in 1970s\cite{Efimov1, Efimov2}.
Theoretically, the Efimov effect can be easily seen in the hyperspherical formalism \cite{Nielsen} where a scale-invariant \( -r^{-2} \) potential emerges at long distances with \( r \) being the hyper radius.
From the renormalization point of view, it corresponds to a limit cycle behavior of the RG flow \cite{Braaten}.
The Efimov effect has been observed in hypernuclei\cite{Lim1}, halo nuclei\cite{Fedorov},
and helium-4 atoms\cite{Lim2, Bruhl}.
In the past two decades, it has attracted renewed interest due to the fast development of experimental techniques of ultracold atomic physics, which offers an ideal platform for the study of such an effect.
Measurement of loss rate of ultracold Bose gases\cite{Kraemer,Gross,Berninger} which directly reflects the effect of underlying Efimov trimers, is in good agreement with theoretical predictions \cite{Braaten}.
These few-body studies further provide valuable input for the understanding of more sophisticated quantum many-body problems.


Although the Efimov effect, or Borromean state in general, is identified as a low-energy/long-range correlation effect, the original non-relativistic(NR) model does not apply to particles that do not have a proper non-relativistic limit, such as the massless case. It is thus appealing to ask whether such exotic few-body states exist in relativistic cases.
Moreover, this question has much realistic meaning in high energy physics. In relativistic heavy ion collisions, the generated extremely hot quark-gluon plasma(QGP) contains huge number of particles and anti-particles. In such a many-body system a remarkable amount of  novel states, which are too rare to be detected in $p+p$ collisions, may be generated. For example, the population of the very rare heavy state $\Xi^+_{cc}$ could be 4 orders of magnitude larger than that in $p+p$ collisions\cite{Zhao:2016ccp}. The light particles, such as the $\pi$ meson which serves as the Goldstone boson corresponding to the chiral symmetry breaking and light quarks which could be treated as chiral fermions in QGP, may form different kinds of novel few-body states, and hence be observed.
Furthermore, for light quarks, even ordinary few-body bound states, such as baryonic states, may be closely related to the famous quark confinement problem. The related hadronization process are still unclear.
Obviously these processes could only be studied in relativistic models. In this work we will focus on 3-body Borromean states because they are more model-independent and more nontrivial than bound states of a 2-body molecule and one particle, which is essentially more like a 2-body problem.

There are a few works concerning the Efimov effect or Borromean states in relativistic bosonic systems\cite{Lindesey1,Lindesey2,Frederico1,Frederico2} by straightforward Mandelstam variables replacement and Bethe-Salpeter equation.
On the other hand, fermions, which are believed to be trivial because of the  Pauli exclusion principle , have attracted much less interest in the relativistic limit. Nevertheless the potential novel states and quark confinement problem motivate us to reconsider the relativistic 3-body problem, especially
the fermionic systems, more seriously. In order to study 2-body and 3-body interactions simultaneously, we will start from  particle--di-particle
models. The di-particle state, in particular the diquark\cite{Rapp:1997zu,Alford:1998mk,Alford:2002rz,Huang:2004ik,Fukushima:2010bq}, which could be viewed as a 2-body bosonic bound state approximately, is not only a useful effective degree of freedom in the model but also has important and realistic application in the modelling of hadronization and phase diagram studies of baryon-rich matter.
In the fermionic case we will more focus on a momentum-dependent 3-body interaction which is motivated by
heavy baryon studies\cite{Ebert:1996ab} with the similar model.

In this report we will adopt an intuitive and self-consistent approach, i.e. the 3-momentum shell renormalization, to study both the bosonic and fermionic systems in the relativistic limit.
For the relativistic and fermionic case the Bethe-Salpeter equation and Dirac equation approach either is too tedious in both analytical and numerical computation\cite{Mu:2012zz} or has difficulties in constructing a self-consistent relativistic spin-dependent 2-body potential\cite{Zhao:2016ccp, Shi:2019tji}.  Therefore the momentum shell renormalization approach is a worthwhile trial of this problem. We will firstly confirm the Efimov effect in the linear bosonic model as Refs.\cite{Kaplan:1996nv,Bedaque:1997qi} with the simplest coupling in both NR and relativistic limits. For the fermionic case, the fermion--di-fermion model will be introduced. It will be shown that there is an indicator of a single Borromean state in the relativistic limit while the systme becomes trivial in the non-relativitic limit as expected.

\begin{figure}[!hbt]
\begin{center}
\includegraphics[scale=0.45]{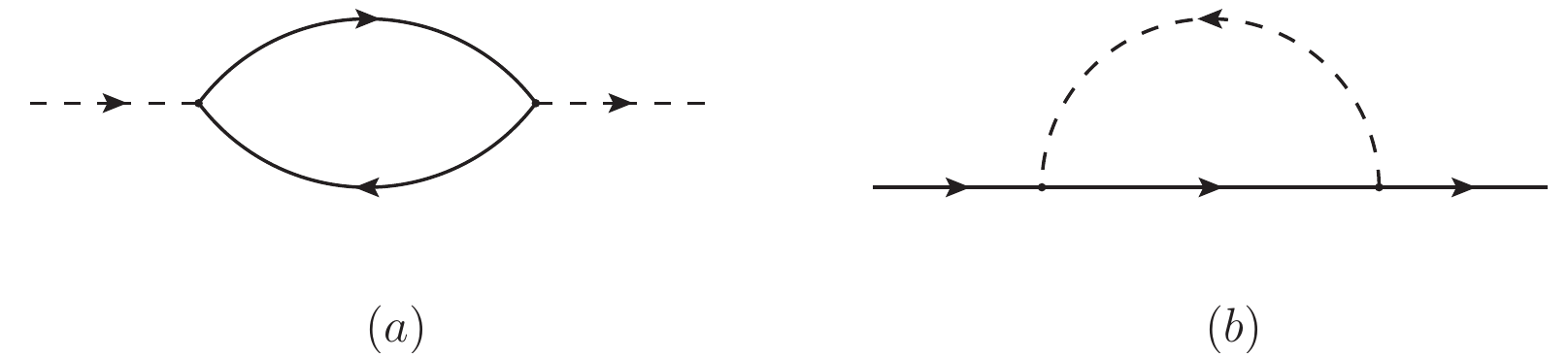}
\caption{1-loop corrections to propagators of $\Delta$ (a) and $\phi$ (b).}
\label{fig_1}
\end{center}
\end{figure}

{\it Bosonic system.--- }
We adopt a boson-di--boson model to study the three-body problem. It is much easier to deal with than the fermionic one because there is no Dirac structure in vertices. The Lagrangian density reads
\begin{eqnarray}
\mathcal{L}&&=\partial^\mu\phi^\dagger\partial_\mu\phi +m^2\phi^2+\partial^\mu\Delta^\dagger\partial_\mu\Delta +M^2\Delta^2\nonumber\\
&&+h (\Delta^\dagger \phi^2+\Delta \phi^{\dagger 2})+g \Delta^2\phi^2,
\end{eqnarray}
where $\Delta$ and $\phi$ are both bosonic fields. Obviously the $\Delta$ field could be viewed as a two-$\phi$ composite field
because of the field equation $\Delta=h \phi^2$ in static limit. We will focus on poles of the 4-point vertex which represents potential
bound states in the $3\phi\rightarrow 3\phi$ scattering process. In order to obtain the flow equation of $g$ we follow the procedure in Refs.\cite{Nishida:2007de, Moroz:2013kf} by computing the perturbative corrections to the 2, 3 and 4-point vertices with the 3-momentum shell renormalization rather than the usual 4-momentum integration in the high energy physics.
In this work we will treat field masses as free parameters and tune them to achieve two energy-scale limits.
Because of charge conservation there is no one-loop correction to the 3-point vertex. Hence the flow equation of $h$ is given by the wave function renormalization of $\Delta$ and $\phi$ fields in Fig.\ref{fig_1} at leading order, i.e. $\partial_s h=h \partial_s Z_\phi+\frac{1}{2}h\partial_s Z_\Delta$. The arrows of propagators represent the charge direction. In non-relatvistic models there is no diagram (b) because of the absence of anti-particles. Here we will see that in NR limit the contribution
of diagram (b) is suppressed by mass.

\begin{figure}[!hbt]
\begin{center}
\includegraphics[scale=0.45]{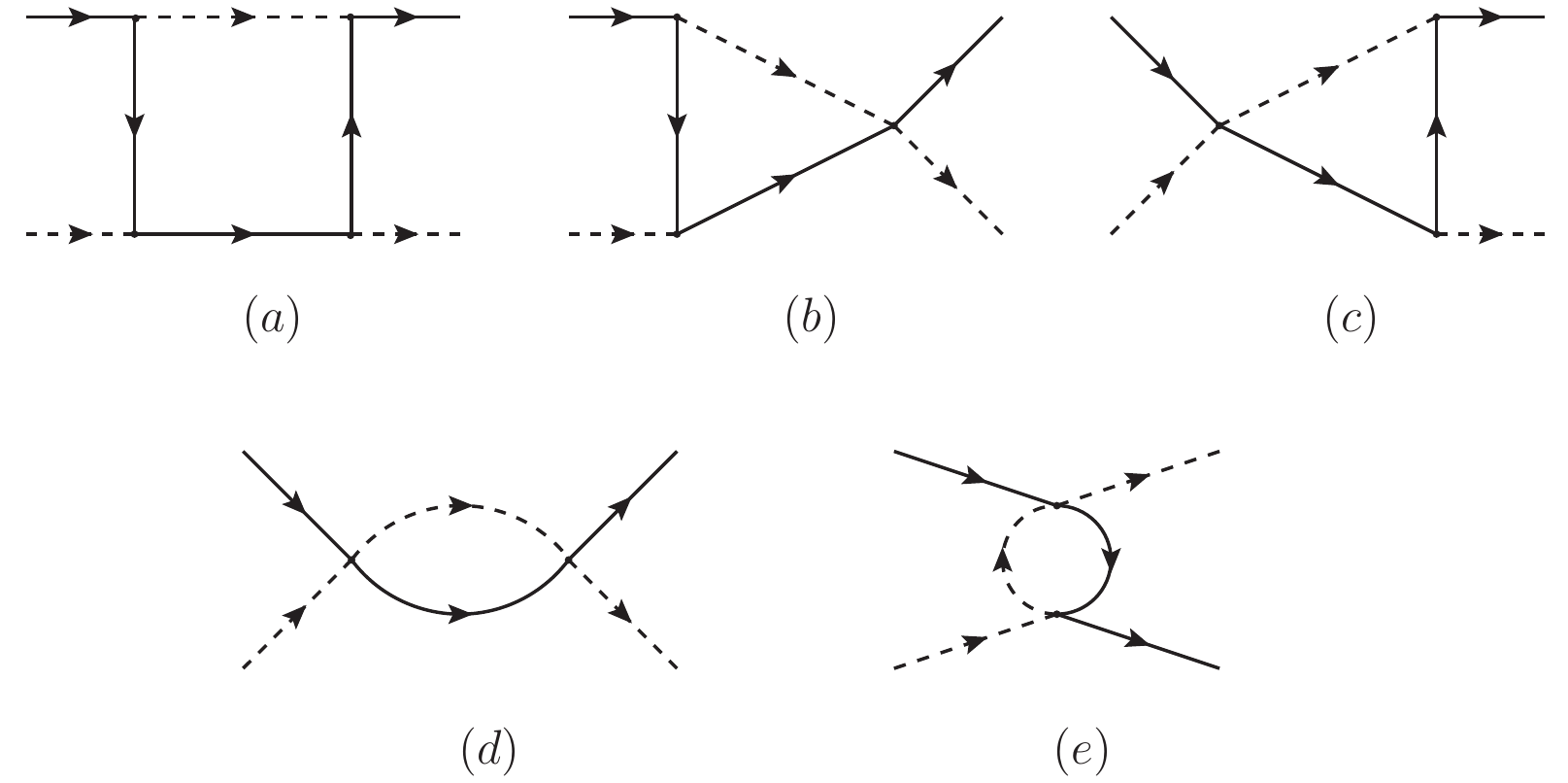}
\caption{1-loop contributions to the renormalization of $g$.}
\label{fig_2}
\end{center}
\end{figure}

Flow equations of wave function renormalization $Z_\Delta$ and $Z_\phi$ and vertices in NR limit,
i.e. $\Lambda \ll m$ and $E_k\simeq m+k^2/(2m)$, are calculated as follows
\begin{eqnarray}
&&\partial_s Z_\Delta=-\frac{2}{32\pi^2}\frac{h^2}{m\Lambda},\ \partial_s Z_\phi=-\frac{2}{32\pi^2}\frac{h^2}{\Lambda^2},\\
&&\partial_s g=\frac{13}{48\pi^2}\frac{g h^2}{m\Lambda}-\frac{1}{3\pi^2}\frac{h^4}{m\Lambda^3}-\frac{1}{12\pi^2}\frac{g^2\Lambda}{m},
\label{l1}
\end{eqnarray}
where $\Lambda=\Lambda_0 e^{-s}$ is the running 3-momentum cutoff. We consider 1-loop corrections to the 4-point vertex coupling $g$ as presented in Fig.\ref{fig_2}. In the NR case the diagram (e) is negligible at the leading order. In the computation, the mass of the two-body field $\Delta$ is chosen as $M=2m$. This is justified by the fact that $h^2=16\pi^2 m\Lambda$ approaches zero in the infrared limit $\Lambda\rightarrow 0$(large $s$), which means that in this limit the two-body state is quite loosely bound. Making use of the result of $h$ in this limit,
the flow equation of coupling $g$ gives the following solution,
\begin{eqnarray}
&&g=\frac{4\pi^2m}{\Lambda}[-\sqrt{39}Tan(\frac{\sqrt{39}}{3}s+c_0)+5].
\end{eqnarray}
Obviously the ratio between binding energies of two neighboring Efimov states is given by $e^{2\delta s}$ with the step of singularities
$\delta s \approx 0.48 \pi$ for this simple model. Besides the expected Efimov-like behavior,
we find that the dimensionless 4-vertex coupling $g$ behaves as a dimensionful quantity scaling with
$\Lambda^{-1}$ in the unit of mass. The same results could be obtained in a explicit NR model by replacing
propagators with $(k_0-{\vec{k}}^2/(2m)+i \eta)^{-1}$.

In relativistic limit, the diagrammatic representations of 1-loop corrections are the same as those in Fig.\ref{fig_1} and Fig.\ref{fig_2}. The difference is that in the limit $\Lambda\gg m$, the diagram (b) in Fig.\ref{fig_1} and diagram (e) in Fig.\ref{fig_2}, due to anti-particles, will be at the same order as the others, and hence contribute to corrections of vertices equally. Taking this anti-particle contribution and relativistic dispersion relation $E_k\simeq k+m^2/(2k)$ into account, the flow equations are

\begin{eqnarray}
&&\partial_s Z_\phi=-\frac{4}{32\pi^2}\frac{h^2}{\Lambda^2},\ \partial_s Z_\Delta=-\frac{2}{32\pi^2}\frac{h^2}{\Lambda^2},\\
&&\partial_s g=\frac{9}{16\pi^2}\frac{g h^2}{\Lambda^2}-\frac{5}{4\pi^2}\frac{h^4}{\Lambda^4}-\frac{g^2}{4\pi^2}.
\label{l2}
\end{eqnarray}
Again we take $M=2m$ in the computation because $h^2=\frac{32}{5}\pi^2\Lambda^2$ tends to vanish in IR limit. Since there is no mass dependence in these equations, their solution would not be changed even in the zero mass limit. And as expected, the wave function renormalization $Z_\phi$ is at the same order as $Z_\Delta$. This antiparticle contribution also appears in the $g^2$ term of Eq.\ref{l2}. In the infrared limit, namely small $\Lambda$ and large $s$, solutions read
\begin{eqnarray}
&&g=-\frac{4\sqrt{239}}{5}\pi^2Tan(\frac{\sqrt{239}}{5}s+c_0)+\frac{36}{5}\pi^2.
\end{eqnarray}
Although the $\Lambda$ dependence of $h$ and $g$ has been changed because of the relativistic dispersion relation, the structure of
solution is qualitatively the same. We get the Efimov-like series of poles with the energy step reduced to $\delta s\simeq 0.323\pi$.

{\it Fermionic system.--- }
As a reducible 4-dimension representation, the Dirac spinor could be studied more straightforwardly and self-consistently in the relativistic quantum field theory. Although it is well-known that the simplest Efimov states does not exist in the NR fermionic system because of the Pauli exclusion principle, we will still examine its Dirac structure, i.e. spin-spin interaction, in detail to find whether there is any non-trivial, such as Borromean,
3-body state. Similar to the bosonic case, we introduce a fermion--di-fermion model to study this problem,
\begin{eqnarray}
\mathcal{L}&&=\bar{\psi}(i \slashed{\partial}-m)\psi+\partial^\mu\Delta^\dagger\partial_\mu\Delta +M^2\Delta^2\nonumber\\
&&-i h (\Delta^\dagger {\bar\psi}_c\gamma^5\psi+\Delta \bar\psi \gamma^5 \psi_c),
\end{eqnarray}
where $\psi_c=C\bar\psi^T=i\gamma^2\gamma^0\bar\psi^T$. This model is motivated by the so-call quark-diquark model in high energy nuclear physics for studies of the quark matter. Here we choose the one-flavor and one-color quark field for simplicity, since the static isospin and color charge will not change momentum dependence of the coupling constants which play crucial roles in the structure of the 4-point vertex flow equation. Effectively the $\psi+\psi_c\rightarrow \Delta$ would generate two kinds of 4-point interactions, i.e. the s-wave $g_1 \Delta^2\bar\psi \psi$ and the p-wave $g_2 \Delta^2\bar\psi {i \slashed \partial}\psi$ because of the Dirac structure. This could be easily checked by straightforward perturbative computation.
Although in principle the s-wave part is supposed to vanish in the one-flavor and one-color scenario because of the Pauli exclusive principle, we still keep it for the following two reasons. First, the Pauli exclusive principle could be detoured by introducing more static color or flavor numbers which will bring no changes to the momentum dependence of the coupling constants up to some symmetric factors. Second, the di-fermion is treated as a fundamental field in this model. As a result this will not forbid the s-wave 3-body interaction technically. It therefore gives us a chance to study its flow equation qualitatively in this simple model. For simplicity we will neglect mixing processes, i.e. $g_1g_2$ terms, and calculate corrections to the flow of the two 4-point vertices separately.

Contributing diagrams are the same as those in Figs.\ref{fig_1} and \ref{fig_2} if we replace the solid propagators with fermionic ones.
In the NR case for the s-wave coupling we get the flow equations as
\begin{eqnarray}
&&\partial_s Z_\Delta=-\frac{m h^2}{4\pi^2\Lambda},\ \partial_s Z_\psi=\frac{h^2 \Lambda^3}{18\pi^2 m^3},\\
&&\partial_s g_1=-\frac{5m h^2 g_1}{12\pi^2 \Lambda}-\frac{m^2 h^4}{12\pi^2\Lambda^3}-\frac{g_1^2 \Lambda}{12\pi^2}.
\end{eqnarray}
Similar to the bosonic case, the $Z_\psi$ term, due to the anti-fermion, is suppressed by the mass as $m^{-3}$. By neglecting the anti-fermion's contribution safely and taking the small $\Lambda$ and large $m$ limit we obtain solutions as
\begin{eqnarray}
&&g_1=\frac{2\sqrt{21}\pi^2}{\Lambda}[Tanh(\frac{\sqrt{21}s}{6}+c_0)-\frac{5}{\sqrt{21}}],
\end{eqnarray}
where $M$ has been set as $2m$ since two-body coupling $h^{2}=4\pi^2\frac{\Lambda}{m}$ approaches to zero in the small $\Lambda$ limit. When the system contains only s-wave coupling, there is no singularity along $s$ for the hyperbolic tangent function, which means no 3-body state appears. This is the well-known result in the NR fermionic system. As a byproduct we also find that the 3-point vertex $h$ is suppressed by $m$ as well which agrees with the $\Delta \psi \psi$ term's vanishing in the NR model because of the fermionic anti-exchange property. In contrast the bosonic result
is enhanced by the mass.

Although the p-wave vertex has a different momentum
dependence, it does not give us more surprise either. The flow equation of $g_2$ reads
\begin{eqnarray}
\partial_s g_2=-\frac{7m h^2 g_2}{12\pi^2\Lambda}-\frac{m h^4}{12\pi^2\Lambda^3}-\frac{m g_2^2 \Lambda}{3\pi^2}.
\end{eqnarray}
Only the $g_2 h^2$ and $g_2^2$ terms are modified.  We get the similar hyperbolic tangent solution as
\begin{eqnarray}
g_2=-\frac{\pi^2}{m \Lambda}[\sqrt{21}Tanh(\frac{\sqrt{21}}{3}s+c_0)-5].
\end{eqnarray}

\begin{figure}[!hbt]
\begin{center}
\includegraphics[scale=0.4]{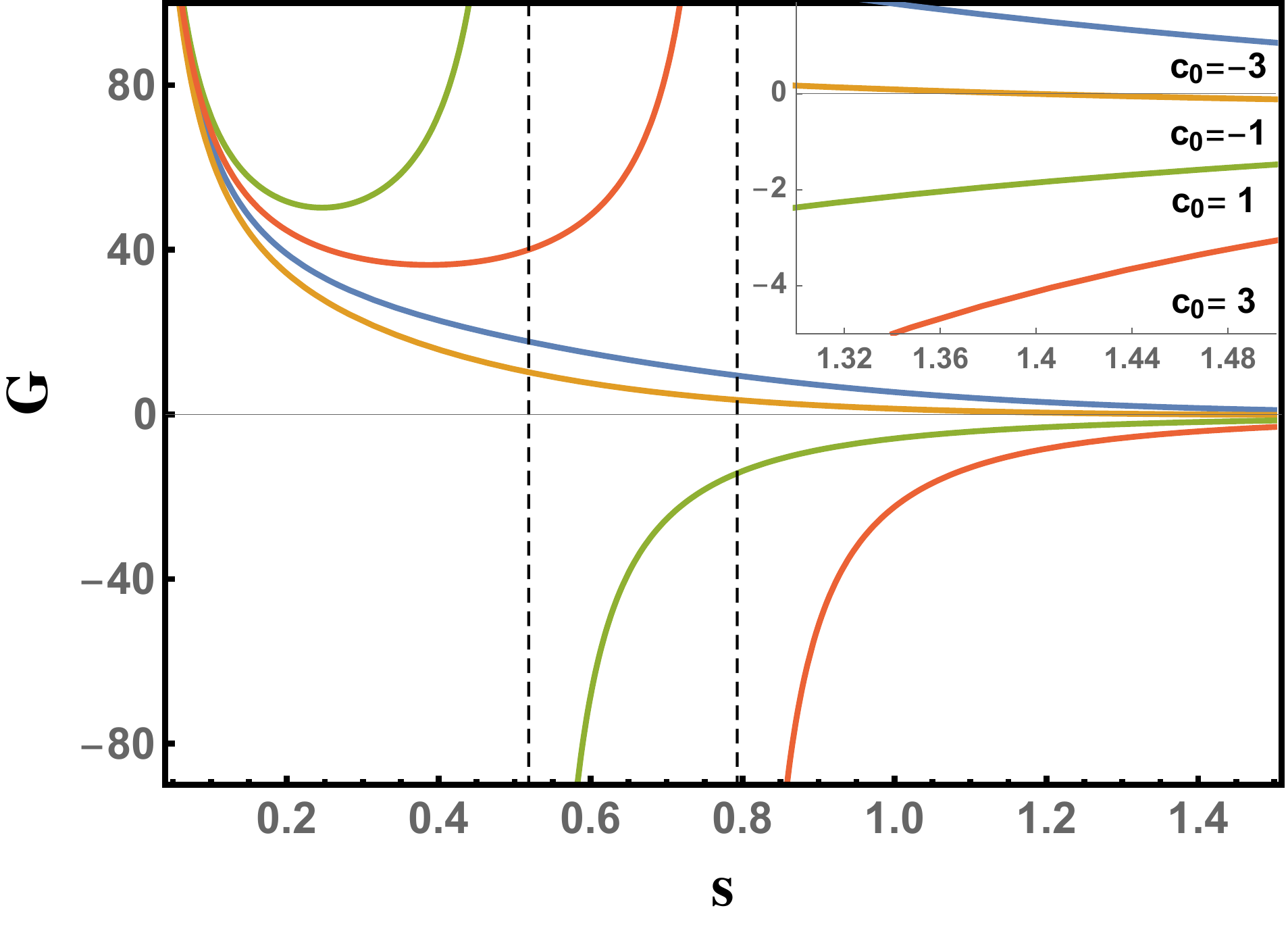}
\caption{The flow of $G$ with integral constants $c_0=-3, -1, 1, 3$.}
\label{fig_3}
\end{center}
\end{figure}

In the relativistic case, i.e. $m\ll\Lambda$, the s-wave case is trivial. The $g_1^2$ and $h^4$ terms are proportional to $m$, therefore the flow of $g_1$ should be governed by $g_1 h^2$ term and results in a trivial solution $g_1\sim s^{-1}$. In the following we focus on the p-wave interaction whose flow equations are
\begin{eqnarray}
&&\partial_s Z_\Delta=-\frac{h^2}{4\pi^2},\ \partial_s Z_\psi=-\frac{h^2}{4\pi^2},\\
&&\partial_s g_2=-\frac{7h^2 g_2}{8\pi^2}-\frac{5h^4}{32\pi^2\Lambda^2}+\frac{3 g_2^2 \Lambda^2}{2\pi^2}.
\end{eqnarray}
Noting that there are no mass dependence, thus the solution would not be altered even in the chiral limit $m=0$. In small $\Lambda$ limit we obtain the two-body coupling goes to zero as $h^2=4\pi^2/(3s)$. And $g_2$ should satisfy
\begin{eqnarray}
\partial_s G=-2G -\frac{5\pi^2}{18s^2}-\frac{7}{6s}G+\frac{3}{2\pi^2}G^2,
\end{eqnarray}
where $G=\Lambda^2 g_2$. The solution goes to $-e^{-2s}$ at large $s$ and converges to $\pi^2(\sqrt{61}+1)s^{-1}/18$ at small $s$.
As a typical Riccati equation, it is usually linearized with an auxiliary function $u(s)$ as
$G(s)=-\frac{2\pi^2}{3}\frac{\partial_s u}{u}$.
And the corresponding solution is
\begin{eqnarray}
u(s)=c_1e^{2s}s^{\beta-\alpha}[c_0 U_{\alpha}^{\beta}(2s)+ L_{-\alpha}^{\beta-1}(2s)]
\end{eqnarray}
where $\alpha=(13+\sqrt{61})/12$, $\beta=\sqrt{61}/6+1$, $U_\alpha^\beta$ the Tricomi confluent hypergeometric function and $L_\alpha^\beta$ the generalized Laguerre polynomial. Obviously the zeros of $u(s)$ will generate singularities of $G(s)$. When the integral constant $c_0>0$, there is an isolated zero of $u(s)$ which generates a 1st order pole of $G(s)$ at finite $s$, and the corresponding pole increases with c0.
When $c_0<0$, the pole approaches zero smoothly. The flow of $G$ with different $c_0$ are presented in Fig.\ref{fig_3}.

At UV range(small $s$) all of the lines converge to the same value $\pi^2\beta/(3s)$, while at IR range(large $s$) different $c_0$ corresponds to different values at low energy-scale. However it is not easy to tune the coupling in the deep IR range to get the 3-body Borromean state, since the value differences are quite small in this range.

{\it Summary and Discussions.--- }
With the linear bosonic model we checked the existence of Efimov-like states in bosonic systems. In NR limit we confirmed that firstly contribution from antiparticles is negligible as expected. Secondly, a series of Efimov-like singularities is found in the flow of the 4-point vertex.
Although more detailed computations are needed to reduce the deviation from the standard value of $\delta s \approx \pi$, qualitatively the NR approximation is promising. In the relativistic limit we find the Efimov-like effect still exists but with a smaller energy ratio. Furthermore, the mass dependence disappears in flow equations, which means the conclusion would hold even in massless limit. For the fermionic case the fermion--di-fermion model,
motivated by the quark-diquark model in the quark matter phase diagram and heavy baryon studies, is adopted. Because of the Dirac structure of
fermions there are two kinds of 4-point vertices that should be taken into account, i.e. the s-wave and p-wave ones. In NR limit both vertices are trivial. The 3-vertex is suppressed by $\Lambda/m$ and thus vanish in the large mass limit. This agrees with straightforward computations of NR models.

In the relativistic limit the result is more non-trivial. Depending on the IR value
of the p-wave 4-vertex coupling there may be one isolated pole at finite $s$. In quark matter the quark-diquark coupling originates from the fundamental strong interaction between color charge, so this singularity may indicate a special baryonic state of quarks. As the bosonic case, this singularity has no mass dependence which means it would persist even for the chiral fermions. For both bosonic and fermionic systems,
all of the singularities appear at the IR range where the 3-vertex approaches zero. Therefore they could be identified as Borromean states. Although it is an exciting clue that there may be a single Borromean state in the chiral fermion case, we should admit the detection is actually difficult by considering the full interaction of different kinds of charges, such as color and isospin. The gauge fields may easily destroy the bound state due to the large amplitude of processes involving soft gauge bosons. In order to determine the existence of the state more works are required, such as fitting the IR value of the model and estimating the scattering
amplitude of different processes from a more realistic model.
And we also believe a more definite answer could be obtained by solving the 3-body Dirac equation with a suitable spin dependent 2-body potential, such as that in \cite{Zhao:2016ccp}.

{\bf Acknowledgments.}
The work of this research is supported by the National Natural Science Foundation of China, Grant Nos. 11875002(YJ) and 11804376(SJ), Postdoctoral Innovative Talent Support Program of China, Grant No. BX20190180(ZyW). YJ is also supported by the Zhuobai Program of Beihang University.

%
%
%
%

\end{document}